\RequirePackage{lineno}
\documentclass[aps,prd,superscriptaddress,showpacs]{revtex4}
\usepackage{epsfig}
\usepackage{graphicx}
\usepackage{dcolumn}
\usepackage{amsmath}
\usepackage{subfigure}
\usepackage{xspace}
\usepackage{color}
\usepackage{amssymb}
\usepackage{epsfig}
\usepackage{footnote}
\usepackage{longtable}
\usepackage{fancyhdr}
\usepackage{subfigure}
\usepackage{xspace}
\usepackage{multirow}
\usepackage{comment}

\def \epage {\enlargethispage*{2.0in}}
\def \epage2 {\enlargethispage*{2.0in}}
\def \epage5 {\enlargethispage*{5.0in}}
\def \bc {\begin{center}}
\def \ec {\end{center}}

\def \Del \nabla
\def \del \partial
\def \hbar {\not h}
\def \ppm $\pm $
\def \D0 {D\O}

\def\pizero{\pi ^0}

\def\gbjmet{\gamma bj\met }

\def\deg{^\circ}
\def\degrees{^\circ}

\def\GeVc2{GeV/{c^2}}

\def\nb-1{nb^{-1}}
\def\pb-1{pb^{-1}}
\def\fb-1{fb^{-1}}

\def\ppbar{{\bar p}p}

\def\Z{${\em Z }$}

\def\Z0{{ Z^0}}

\def\95cl{95 \%~C.L.}
\def\95CL{95 \%~C.L.}
\def\r#1 {$^{#1}$}

\def\M4j{M_{4j}}
\def\M12{M_{12}}
\def\M34{M_{34}}

\newcommand{\met}{\mbox{${\rm \not\! E}_{T}$}}

\newcommand{\degs}{\mbox{$^{\circ}$}}

\newcommand{\metvec}{{\not\!\! \vec{E}_T}}
\newcommand{\wgamma}{\ensuremath{{\cal W}_{\gamma}}}
\renewcommand{\gbjmet}{\gamma bj {\not\!\! {E}_T}}

\begin{document}

\title{Search for Anomalous Production of Events with a Photon, Jet, $b$-quark Jet, and Missing Transverse Energy}
\affiliation{Institute of Physics, Academia Sinica, Taipei, Taiwan 11529, Republic of China} 
\affiliation{Argonne National Laboratory, Argonne, Illinois 60439} 
\affiliation{University of Athens, 157 71 Athens, Greece} 
\affiliation{Institut de Fisica d'Altes Energies, Universitat Autonoma de Barcelona, E-08193, Bellaterra (Barcelona), Spain} 
\affiliation{Baylor University, Waco, Texas  76798} 
\affiliation{Istituto Nazionale di Fisica Nucleare Bologna, $^v$University of Bologna, I-40127 Bologna, Italy} 
\affiliation{Brandeis University, Waltham, Massachusetts 02254} 
\affiliation{University of California, Davis, Davis, California  95616} 
\affiliation{University of California, Los Angeles, Los Angeles, California  90024} 
\affiliation{University of California, San Diego, La Jolla, California  92093} 
\affiliation{University of California, Santa Barbara, Santa Barbara, California 93106} 
\affiliation{Instituto de Fisica de Cantabria, CSIC-University of Cantabria, 39005 Santander, Spain} 
\affiliation{Carnegie Mellon University, Pittsburgh, PA  15213} 
\affiliation{Enrico Fermi Institute, University of Chicago, Chicago, Illinois 60637}
\affiliation{Comenius University, 842 48 Bratislava, Slovakia; Institute of Experimental Physics, 040 01 Kosice, Slovakia} 
\affiliation{Joint Institute for Nuclear Research, RU-141980 Dubna, Russia} 
\affiliation{Duke University, Durham, North Carolina  27708} 
\affiliation{Fermi National Accelerator Laboratory, Batavia, Illinois 60510} 
\affiliation{University of Florida, Gainesville, Florida  32611} 
\affiliation{Laboratori Nazionali di Frascati, Istituto Nazionale di Fisica Nucleare, I-00044 Frascati, Italy} 
\affiliation{University of Geneva, CH-1211 Geneva 4, Switzerland} 
\affiliation{Glasgow University, Glasgow G12 8QQ, United Kingdom} 
\affiliation{Harvard University, Cambridge, Massachusetts 02138} 
\affiliation{Division of High Energy Physics, Department of Physics, University of Helsinki and Helsinki Institute of Physics, FIN-00014, Helsinki, Finland} 
\affiliation{University of Illinois, Urbana, Illinois 61801} 
\affiliation{The Johns Hopkins University, Baltimore, Maryland 21218} 
\affiliation{Institut f\"{u}r Experimentelle Kernphysik, Universit\"{a}t Karlsruhe, 76128 Karlsruhe, Germany} 
\affiliation{Center for High Energy Physics: Kyungpook National University, Daegu 702-701, Korea; Seoul National University, Seoul 151-742, Korea; Sungkyunkwan University, Suwon 440-746, Korea; Korea Institute of Science and Technology Information, Daejeon, 305-806, Korea; Chonnam National University, Gwangju, 500-757, Korea} 
\affiliation{Ernest Orlando Lawrence Berkeley National Laboratory, Berkeley, California 94720} 
\affiliation{University of Liverpool, Liverpool L69 7ZE, United Kingdom} 
\affiliation{University College London, London WC1E 6BT, United Kingdom} 
\affiliation{Centro de Investigaciones Energeticas Medioambientales y Tecnologicas, E-28040 Madrid, Spain} 
\affiliation{Massachusetts Institute of Technology, Cambridge, Massachusetts  02139} 
\affiliation{Institute of Particle Physics: McGill University, Montr\'{e}al, Qu\'{e}bec, Canada H3A~2T8; Simon Fraser University, Burnaby, British Columbia, Canada V5A~1S6; University of Toronto, Toronto, Ontario, Canada M5S~1A7; and TRIUMF, Vancouver, British Columbia, Canada V6T~2A3} 
\affiliation{University of Michigan, Ann Arbor, Michigan 48109} 
\affiliation{Michigan State University, East Lansing, Michigan  48824}
\affiliation{Institution for Theoretical and Experimental Physics, ITEP, Moscow 117259, Russia} 
\affiliation{University of New Mexico, Albuquerque, New Mexico 87131} 
\affiliation{Northwestern University, Evanston, Illinois  60208} 
\affiliation{The Ohio State University, Columbus, Ohio  43210} 
\affiliation{Okayama University, Okayama 700-8530, Japan} 
\affiliation{Osaka City University, Osaka 588, Japan} 
\affiliation{University of Oxford, Oxford OX1 3RH, United Kingdom} 
\affiliation{Istituto Nazionale di Fisica Nucleare, Sezione di Padova-Trento, $^w$University of Padova, I-35131 Padova, Italy} 
\affiliation{LPNHE, Universite Pierre et Marie Curie/IN2P3-CNRS, UMR7585, Paris, F-75252 France} 
\affiliation{University of Pennsylvania, Philadelphia, Pennsylvania 19104}
\affiliation{Istituto Nazionale di Fisica Nucleare Pisa, $^x$University of Pisa, $^y$University of Siena and $^z$Scuola Normale Superiore, I-56127 Pisa, Italy} 
\affiliation{University of Pittsburgh, Pittsburgh, Pennsylvania 15260} 
\affiliation{Purdue University, West Lafayette, Indiana 47907} 
\affiliation{University of Rochester, Rochester, New York 14627} 
\affiliation{The Rockefeller University, New York, New York 10021} 
\affiliation{Istituto Nazionale di Fisica Nucleare, Sezione di Roma 1, $^{aa}$Sapienza Universit\`{a} di Roma, I-00185 Roma, Italy} 

\affiliation{Rutgers University, Piscataway, New Jersey 08855} 
\affiliation{Texas A\&M University, College Station, Texas 77843} 
\affiliation{Istituto Nazionale di Fisica Nucleare Trieste/Udine, I-34100 Trieste, $^{bb}$University of Trieste/Udine, I-33100 Udine, Italy} 
\affiliation{University of Tsukuba, Tsukuba, Ibaraki 305, Japan} 
\affiliation{Tufts University, Medford, Massachusetts 02155} 
\affiliation{Waseda University, Tokyo 169, Japan} 
\affiliation{Wayne State University, Detroit, Michigan  48201} 
\affiliation{University of Wisconsin, Madison, Wisconsin 53706} 
\affiliation{Yale University, New Haven, Connecticut 06520} 
\author{T.~Aaltonen}
\affiliation{Division of High Energy Physics, Department of Physics, University of Helsinki and Helsinki Institute of Physics, FIN-00014, Helsinki, Finland}
\author{J.~Adelman}
\affiliation{Enrico Fermi Institute, University of Chicago, Chicago, Illinois 60637}
\author{T.~Akimoto}
\affiliation{University of Tsukuba, Tsukuba, Ibaraki 305, Japan}
\author{B.~\'{A}lvarez~Gonz\'{a}lez$^q$}
\affiliation{Instituto de Fisica de Cantabria, CSIC-University of Cantabria, 39005 Santander, Spain}
\author{S.~Amerio$^w$}
\affiliation{Istituto Nazionale di Fisica Nucleare, Sezione di Padova-Trento, $^w$University of Padova, I-35131 Padova, Italy} 

\author{D.~Amidei}
\affiliation{University of Michigan, Ann Arbor, Michigan 48109}
\author{A.~Anastassov}
\affiliation{Northwestern University, Evanston, Illinois  60208}
\author{A.~Annovi}
\affiliation{Laboratori Nazionali di Frascati, Istituto Nazionale di Fisica Nucleare, I-00044 Frascati, Italy}
\author{J.~Antos}
\affiliation{Comenius University, 842 48 Bratislava, Slovakia; Institute of Experimental Physics, 040 01 Kosice, Slovakia}
\author{G.~Apollinari}
\affiliation{Fermi National Accelerator Laboratory, Batavia, Illinois 60510}
% ...

\collaboration{CDF Collaboration\footnote{With visitors from $^a$University of Massachusetts Amherst, Amherst, Massachusetts 01003,
$^b$Universiteit Antwerpen, B-2610 Antwerp, Belgium, 
$^c$University of Bristol, Bristol BS8 1TL, United Kingdom,
$^d$Chinese Academy of Sciences, Beijing 100864, China, 
$^e$Istituto Nazionale di Fisica Nucleare, Sezione di Cagliari, 09042 Monserrato (Cagliari), Italy,
$^f$University of California Irvine, Irvine, CA  92697, 
$^g$University of California Santa Cruz, Santa Cruz, CA  95064, 
$^h$Cornell University, Ithaca, NY  14853, 
$^i$University of Cyprus, Nicosia CY-1678, Cyprus, 
$^j$University College Dublin, Dublin 4, Ireland,
$^k$University of Edinburgh, Edinburgh EH9 3JZ, United Kingdom, 
$^l$Universidad Iberoamericana, Mexico D.F., Mexico,
$^m$Queen Mary, University of London, London, E1 4NS, England,
$^n$University of Manchester, Manchester M13 9PL, England, 
$^o$Nagasaki Institute of Applied Science, Nagasaki, Japan, 
$^p$University of Notre Dame, Notre Dame, IN 46556,
$^q$University de Oviedo, E-33007 Oviedo, Spain, 
$^r$Texas Tech University, Lubbock, TX  79409, 
$^s$IFIC(CSIC-Universitat de Valencia), 46071 Valencia, Spain,
$^t$University of Virginia, Charlottesville, VA  22904,
$^u$Bergische Universit\"at Wuppertal, 42097 Wuppertal, Germany,
$^{cc}$On leave from J.~Stefan Institute, Ljubljana, Slovenia, 
}}
\noaffiliation

\begin{abstract}
We present a signature-based search for anomalous production of
events containing a photon, two jets, of which at least one is
identified as originating from a $b$ quark, and missing transverse
energy ($\met$).  The search uses data
corresponding to $2.0$ $\mathrm{fb}^{-1}$ of integrated luminosity
from $p\bar{p}$ collisions at a center-of-mass energy of
$\sqrt{s}=1.96$ TeV, collected 
with the CDF II detector at the Fermilab Tevatron. From 6,697,466
\ events with a photon candidate with
 transverse energy $E_T> 25$ GeV, we find 617 events with $\met> 25$ GeV and
two or more jets with $E_T> 15$ GeV, at least one identified as
originating from a $b$ quark, versus an expectation
of $607\pm 113$ events. Increasing the requirement on $\met$ to 50
GeV, we find 28 events versus an expectation of
$30\pm11$ events.  We find no indications of non-standard-model phenomena.
\end{abstract}

\maketitle

\section{Introduction}
\label{sec:introduction}
Within the standard model of elementary 
particle physics (SM) there are six flavors of quarks, six flavors of
leptons, and four vector gauge bosons, with a hierarchy of couplings
and masses. The Fermilab Tevatron, with a center-of-mass energy of
1.96 TeV, can produce all of the known quarks and vector bosons. 
Over the course of years of data-taking using the CDF detector~\cite{CDF}, 
we have developed a suite of largely data-driven methods by which we 
estimate the efficiencies and backgrounds associated with the identification 
of charged leptons, heavy flavor quarks ($b$ or $c$ quark), electroweak gauge 
bosons (photon, $W^{\pm}$, and $Z^0$), and the presence of neutrinos, 
identified generically by missing transverse energy ($\met$)~\cite{EtPt}. 
The ability to identify these `objects' in events and to estimate their 
efficiencies and backgrounds has led to the development of 
signature-based searches at the Tevatron, in which one defines {\it a priori} 
the objects an event is required to contain, and then compares observations to
expectations~\cite{Toback_all,Berryhill_all,onyisi_gamma_met,andrei_gamma_lepton,toback_delayed_photons,diphoton_X,Knuteson}.
The model tested in these searches is the SM, which is
predictive and falsifiable; any deviation from the SM predictions
would be a signal of new phenomena. The advantage of this strategy is
that only once such a signal has been established would the investment
be made in generating detailed predictions of the many possible models
for the new phenomena.

We describe here a search for new physics in the inclusive
$\gbjmet$ channel using $2.0\pm0.1$ $\mathrm{fb}^{-1}$ of integrated
luminosity at a center-of-mass energy of $\sqrt{s}=1.96$ TeV, collected between
February 2002 and May 2007.  A similar search was originally performed in
Run I using $\approx 85$ $\mathrm{pb}^{-1}$ of integrated luminosity~\cite{RunI_gammabjmet}. 
%
%
Our search in Run II is part of a broad effort at CDF to study rare
event signatures involving photons for any non-SM
sources~\cite{andrei_gamma_lepton,toback_delayed_photons,diphoton_X}.
The SM processes, either with a radiated
photon or where the charged lepton is misidentified as a photon, are expected
to contribute $\approx 2\%$ ($t\bar{t}\rightarrow \ell\bar{\nu}jjb\bar{b}$) 
and $<1\%$ ($Wb\bar{b}\rightarrow\ell\bar{\nu}b\bar{b}$ and
$Zb\bar{b}\rightarrow\nu\bar{\nu}b\bar{b}$) 
to the measured rate~\cite{silly_2percent}.
Because the SM contributions to the $\gbjmet$ final state are highly
suppressed, for an ideal detector the signature provides an excellent
place to look for new phenomena.  In reality, we expect additional
events from processes such as $\gamma$ + jets and $b\bar{b}$
production in which mismeasurements of the jet energy induce $\met$.

%
The outline of the paper is as follows.  Section~\ref{sec:detector}
briefly describes the CDF II detector.  The selection of events with
photons, jets, jets from a heavy flavor quark ($b$ or $c$ quark), and missing 
transverse energy is described in Sec.~\ref{sec:evselect}.  The estimation of
backgrounds to the search sample is presented in Sec.~\ref{sec:bg}.
Section~\ref{sec:systErrs} describes the sources and estimates of
systematic uncertainties on the numbers of events from backgrounds.
The results of the search, including the effect of additional
selection criteria,
are presented in 
Sec.~\ref{sec:results}.  Section~\ref{sec:conclusions} presents the conclusions.

\section{The CDF II Detector}
\label{sec:detector}
The CDF II detector is a cylindrically-symmetric
spectrometer designed to study $\ppbar$ collisions at the Fermilab
Tevatron. The detector has been extensively described in detail in
the literature~\cite{CDF}.  Here we briefly describe the
detector subsystems relevant for the analysis.

Tracking systems are used to measure the momenta of charged particles, to 
reconstruct primary and secondary vertices, and to trigger
 on and identify leptons with large transverse momentum~\cite{EtPt}. 
Silicon strip detectors~\cite{SVX} and the central 
outer tracker (COT)~\cite{COT} are contained in a superconducting solenoid 
that generates a magnetic field of 1.4~T. 
The silicon strip system provides up to 8 measurements in the 
$r-\phi$ and $r-z$ views~\cite{EtPt} and covers the track reconstruction in 
the region $|\eta| <$ 2. 
The COT is an open-cell drift chamber that makes up to 96 
measurements along the track of each charged particle in the
region $|\eta|<1$. Sense wires are arranged in 8 alternating axial
and $\pm 2\degs$ stereo super-layers. 
The resolution in $p_T$, $\sigma_{p_T}/p_T$, is 
$\approx 0.0015\;p_T\cdot{\rm GeV}^{-1}\cdot c$ for tracks with only COT measurements, and 
$\approx 0.0007\;p_T\cdot{\rm GeV}^{-1}\cdot c$ for tracks with both the silicon and COT 
measurements.

Calorimeters are segmented with towers arranged in a projective geometry. 
Each tower consists of an electromagnetic and a hadronic
compartment~\cite{cem_resolution,cha,cal_upgrade}, covering the central
region, $|\eta|< 1.1$ and the `end plug' region, 
$1.1<|\eta|<3.6$.  The central electrogmagnetic calorimeter (CEM) and central hadronic calorimeter (CHA)
are in the central region while the plug electromagnetic calorimeter (PEM) and plug hadronic
calorimeter (PHA) are in the `end plug' region.
In this analysis, a high-$E_T$ photon is
required to be identified in the central region, where the CEM has a
segmentation of  $15^{\degrees}$ in $\phi$ and $\approx 0.1$ in
$\eta$~\cite{CDF},
and an $E_T$ resolution of $\sigma(E_T)/E_T \approx
13.5\%/\sqrt{E_T/\rm{GeV}}\oplus 2\%$~\cite{cem_resolution}.  We further
require a high-$E_T$ jet to be identified in the central region, where
the jet energy resolution is approximately $\sigma \approx 0.1\cdot E_T
({\rm{GeV}}) \oplus 1.0~\rm{GeV}$~\cite{jet_resolution}.
Two additional systems in the central region with finer spatial
resolution are used for photon identification in this analysis.  The
central strip system, CES, uses a multi-wire proportional chamber
with 1.67- and 2.01-cm-wide cathode strips and a wire spacing of 1.45
cm to make profile measurements of electromagnetic showers at a depth
of 6 radiation lengths (approximately shower maximum).  The central
preshower detector, CPR, located just outside the solenoid coil on
the front face of the CEM, separates single photons from the photon
pairs from $\pizero$ and $\eta^0$ decays on a statistical basis, as
described in Sec.~\ref{sec:falsephoton}.  In 2005 the CPR was upgraded 
from the Run I configuration of wire proportional chambers, similar to 
those used in the CES, to a fast scintillator system with a segmentation 
of 12.5~cm in $\phi$ and 12.5~cm in $z$~\cite{cal_upgrade}.  The finer 
segmentation in $z$ reduces the probability of a random hit from the 
underlying event and multiple interactions by a factor of four, thereby 
improving the performance of the preshower detector in higher luminosity 
beam conditions.

Muons are identified using the central muon systems~\cite{muon_systems}: 
CMU and CMP for the pseudorapidity region of $|\eta|<0.6$, and
CMX for the pseudorapidity region of $0.6<|\eta|<1.0$. The CMU system uses 
four layers of planar drift chambers to detect muons with $p_T > 1.4$ GeV/$c$.
 The CMP system consists of an additional four layers
of planar drift chambers located behind 0.6 m of steel outside the
magnetic return yoke, and detects muons with $p_T > 2.2$ GeV/$c$. 
The CMX system detects muons with $p_T > 1.4$ GeV/$c$ with four to
eight layers of drift chambers, depending on the direction of the muon.

The luminosity is measured using two sets of gas Cerenkov
counters~\cite{CLC}, located in the region $3.7<|\eta|<4.7$. The total
uncertainty on the luminosity is estimated to be 5.9\%, where
4.4\% comes from the acceptance and operation of the luminosity
monitor and 4.0\% from the calculation of the inelastic $\ppbar$
cross-section~\cite{LumiUncertainty}.

A three-level trigger system~\cite{trigger} selects events to be recorded
for further analysis. The first two trigger levels consist of
dedicated fast digital electronics analyzing a subset of the full
detector data.  The third level, applied to the full data of those
events passing the first two levels, consists of a farm of computers
that reconstruct the data and apply selection criteria 
consistent with the subsequent offline event processing.

%


\section{Event Selection}
\label{sec:evselect}
An initial sample of events enhanced with high energy
photons is collected using a trigger that requires a
high energy isolated cluster in the electromagnetic 
calorimeter~\cite{Photon25Trig}.
We require events to have a primary vertex with $|z|<60$ cm.
The offline selection criteria require a central  ($|\eta|<1.1$)
photon with $E_T>25$ GeV, two jets with $|\eta|<2.0$ and $E_T>15$ GeV,
at least one of which is identified as originating from a $b$
quark ($b$-tagged), and missing transverse energy greater than
25 GeV, as described in more detail below. The selection is
inclusive; i.e. we allow extra objects (jets, photons,
leptons) in the events.

The photon is required to satisfy the same identification requirements
as in previous CDF high-$E_T$ photon
analyses~\cite{diphoton_MET}. Namely, the photon candidate is required
to have no associated track with $p_T>1$ GeV, at most one track with
$p_T<1$ GeV pointing at the calorimeter cluster~\cite{calCluster},
good profiles of electromagnetic energy measured in both transverse dimensions
 at shower maximum, and
minimal leakage into the hadron calorimeter~\cite{photonHADleakage}.
Photon candidates identified via these cuts are referred to as ``standard''
photons.  

Jets are reconstructed using the {\sc jetclu} cone
algorithm~\cite{jetclu} with cone radius
$R=\sqrt{\Delta\phi^2+\Delta\eta^2}=0.4$. Starting from seed
locations corresponding to calorimeter towers with $E_T>1$ GeV,
all nearby towers with $E_T>0.1$ GeV are used to search for stable
cones.  To resolve ambiguities with overlapping cones, cones
sharing an energy fraction greater than $0.75$ are merged into a
single jet; otherwise the shared towers are assigned to the
closest jet. We apply a jet energy scale (JES)
correction~\cite{JES} such that the measured $E_T$
is on average equal to the summed $E_T$ of the particles from the
$p\bar{p}$ interaction within the jet cone.  
At least one of the jets must be $b$-tagged using the {\sc secvtx}
algorithm~\cite{SECVTX}, which searches for displaced vertices
using the reconstructed tracks inside the jet cone.

Missing transverse energy $\met$ is calculated~\cite{EtPt} from the calorimeter
tower energies in the region $|\eta|<3.6$. Corrections are
then applied to the $\met$ for (i) the calorimeter
response for identified jets~\cite{JES}, and (ii) the presence of
muons with $p_T>20$ GeV.  We require the corrected $\met$ to be
greater than $25$ GeV and minimize the number of events with
mismeasured $\met$ by requiring the difference in azimuthal angle
between any jet and the $\met$, $\Delta\phi(\mathrm{jet},\met)$, to be
greater than 0.3.  The requirement $\Delta R > 0.4$ is imposed on all
combinations of the photon and the two selected jets, namely $\gamma
j_1$, $\gamma j_2$, and $j_1 j_2$. One of the two jets is the leading
$b$-tagged jet, and the other is the next-to-leading $b$-tagged jet if
one exists, or the leading non-$b$-tagged jet if not.

Table~\ref{tab:evselect} summarizes the event selection.  The final
  $\gbjmet$ sample with $\met>25$ GeV corresponds to one part in
  $10^4$ of inclusive high-$E_T$ photon events.  We will refer to this
  sample of 617 events as the ``search'' sample.

\begin{table}[htb]
\renewcommand{\tabcolsep}{0.2in}
\caption{Summary of the event selection.  The selection of a central photon
includes the requirement of the inclusive photon trigger, the
selections on the $z$-vertex, and $E_T(\gamma)$ as described in
Sec.~\ref{sec:evselect}.  The selection $\Delta R>0.4$ is required
for each pair of $\gamma j_1$, $\gamma j_2$, and $j_1 j_2$.
\label{tab:evselect}}
\begin{center}
\begin{tabular}{lr}
\hline
\hline
\textbf{Cut} & \textbf{Events} \\
\hline
Photon with $E_T > 25$ GeV, $\left|\eta\right| < 1.1$  & 6,697,466 \\
2 Jets with $E_T>15$ GeV, $\left|\eta\right| < 2.0$ & 1,944,962 \\
$\Delta R>0.4$ for $\gamma j_1$, $\gamma j_2$, and $j_1 j_2$ & 1,941,343 \\
$\met\geq25$ GeV                            & 35,463 \\
$\Delta\phi(\mathrm{jet},\met)>0.3$      & 18,128 \\
$\geq1$ {\sc secvtx} $b$-tag                   & 617 \\
\hline
\hline
\end{tabular}
\end{center}
\end{table}

\section{Background Predictions}
\label{sec:bg}

To understand the composition of the search sample of 617 events, we
could in principle use Monte Carlo (MC) simulations to estimate the
absolute numbers of events with real or misidentified photons, and
real or misidentified heavy flavor.  However, this method would
result in large systematic uncertainties on the number of events due
to theoretical uncertainties on the production cross-sections and
difficulties in modeling misidentifications.  We have consequently
developed a data-driven strategy that uses the Monte Carlo judiciously
to minimize systematic uncertainties.

As mentioned in Sec.~\ref{sec:introduction}, SM processes with final state neutrinos are not
expected to contribute significantly to our search region.  We check this expectation by
vetoing events that have any high $p_{T}$ isolated tracks, effectively removing any contribution from processes
involving leptonic decays of vector bosons.  Isolated
tracks are defined as tracks with $p_T>20$ GeV having an isolation
fraction larger than 0.9, where the isolation fraction is defined as
\begin{eqnarray}
\nonumber
f_{iso}=\frac{p_{T}^{\mathrm{track}}}{p_{T}^{\mathrm{track}}+\sum_{i}p_{T}^{i}}.
\end{eqnarray}
Further details of the isolation calculation are given in Ref.~\cite{isodef}.
After the application of the isolated track veto, the observed number of events decreases from 617 to 600.  This
decrease is consistent with the $\approx 3\%$ expectation obtained from Monte 
Carlo simulation~\cite{silly_2percent}.

We define four categories of dominant background
events, all of which have missing transverse energy primarily arising
from energy mismeasurement. We obtain the kinematic shapes and
normalizations of each category separately.  The four categories are
a) \emph{misidentified photons}, referred to as ``misidentified
$\gamma$'', b) \emph{true photon plus light quark jet misidentified as
heavy flavor}, referred to as ``true $\gamma$, misidentified $b$'', c)
\emph{true photon plus true b-quark jet}, referred to as ``$\gamma
b$'', and d) \emph{true photon plus true c-quark jet}, referred to as
``$\gamma c$''.

%


The misidentified $\gamma$ background is estimated from the data
sample itself by using cluster-shape variables from the CES and hit
rates in the CPR.  This technique (the CES/CPR method)
allows the determination of the number of photon candidates in the sample that
are actually misidentified jets as well as the corresponding shapes of
the distributions of kinematic variables ~\cite{cescp2}. We describe
the method in more detail in Sec.~\ref{sec:falsephoton}.

The true $\gamma$ plus misidentified $b$ background is estimated by
first selecting events that pass all cuts except the requirement of a
$b$-tagged jet, resulting in 18128 events (see
Tab.~\ref{tab:evselect}).  For each selected event we then apply the
product of two weights: (i) the true-photon weight determined using
the CES/CPR method, representing the probability that a photon
candidate is a photon, and (ii) the heavy-flavor mistag~\cite{mistag}
rate, which depends on jet $E_T$, jet $\eta$, the number of tracks in
the jet,the number of primary interactions found in the event, and the
$z$ position of the primary interaction with the highest scalar sum
$p_T$ of tracks. The mistag parameterization is the same
as that used in the measurement of the $t\bar{t}$
cross-section~\cite{SECVTX}.  Because the CES/CPR method and the
mistag parametrization provide event-by-event weights, we are able to
determine the shapes of kinematic distributions as well as the number
of events for this background.

We estimate the $\gamma b$ and $\gamma c$ backgrounds by generating 
MC events using {\sc madgraph}~\cite{madgraph} for leading-order 
matrix element processes involving photons, $b$ or $c$ quarks, and additional 
partons. The samples for $\gamma + b~+$ jets and $\gamma + c~+$ jets are 
generated with 1 to 3 jets. 
These samples are then processed with 
{\sc pythia}~\cite{pythia} to incorporate parton showering and hadronization.  
We ensure that we do not double-count events due to the overlap
between jets arising from matrix element partons and jets arising from initial and final state radiation~\cite{CKKW}.
We obtain the overall normalizations of these backgrounds by fitting the 
secondary vertex mass distribution of the tagged jets, $m(SV)$, to templates formed from the mass 
distributions of the expected SM components.  The normalization scheme
is described in Sec.~\ref{sec:svfit}. 

A summary of the background contributions is given in Tab.~\ref{tab:BGresults}.

\subsection{Photon Backgrounds: The CES/CPR Method}
\label{sec:falsephoton} 
For photon candidates with $E_T<35$ GeV, we use the shape of the shower
 profile measured with the CES system to discriminate between
 true single photon events and diphoton final states from decays of
 mesons. We construct a $\chi^2$ discriminant by comparing the
 measured shower profile with that measured in electron test beam
 data~\cite{cescp2}. A single photon has an average probability of
 $\approx78\%$ to satisfy the $\chi^2$ cut, while the background has an
 average probability of $\approx30\%$ to satisfy the $\chi^2$ cut, since
 the shower profile of the two near-by photons from a meson decay is
 measurably wider on average.

Above $35$ GeV, however, the two photons from meson decay coalesce and 
the discrimination power of the shower profile measurement is significantly 
reduced. 
In this $E_T$ range, we use hit rates in the CPR system to discriminate between single 
photons and diphotons from meson decays. A single photon will convert and leave
 a hit in the preshower detector with a probability of $\approx65\%$.  Backgrounds
 that decay into two photons have a hit probability of $\approx85\%$ because the 
probability that neither photon converts is lower than the probability that a 
single photon does not convert.

The difference of probabilities between signal (single photons) and background 
(photon pairs) forms the basis of a statistical method 
which assigns each event a weight for being a true photon (termed
true-photon weight), $\wgamma$, as described in
Ref.~\cite{cescp2}. The weight is defined as: 
\begin{eqnarray}
 \label{eq:cescpr}
  \wgamma = \frac{\delta_\mathrm{candidate}-\epsilon_\mathrm{bkg}}
	{\epsilon_\mathrm{sig}-\epsilon_\mathrm{bkg}},
\end{eqnarray}
where $\epsilon_\mathrm{sig}$ and $\epsilon_\mathrm{bkg}$ are the
respective probabilities for a true photon and misidentified photon
to satisfy a CES $\chi^2$ cut or to leave a hit in the CPR, and
$\delta_\mathrm{candidate}$ is either zero or one depending on whether
the observed candidate satisfies these CES and CPR conditions.  The
values of $\epsilon_\mathrm{sig}$ and $\epsilon_\mathrm{bkg}$ are
determined using control data samples~\cite{eiko_endnote} and are parametrized as a function
of the energy of the photon candidate, the angle of incidence, and the
number of primary interactions found in the event.  The
misidentified-photon weight is \(1-\wgamma\). We estimate the
misidentified $\gamma$ background by summing up the
misidentified-photon weights of the 617 candidate events ($\gbjmet$) to
obtain 115 events with a statistical uncertainty of 49 events.
We estimate the true $\gamma$, misidentified $b$ background by
summing up the products of true-photon weights and heavy-flavor mistag
rates of the 18128 events before the $b$-tagging selection ($\gamma jj\met$) to obtain 141
events with a statistical uncertainty of 6 events. The calculation of the systematic uncertainty on these
expectations is given in Sec.~\ref{sec:systErrs}.

\subsection{Heavy-Flavor Normalization}
\label{sec:svfit}

The invariant mass of the tracks that form a secondary vertex can be
used to discriminate between the bottom, charm, and light partons that
compose a sample.  We use this discriminating variable to normalize the
contributions of the $\gamma b$ and $\gamma c$ backgrounds by
fitting the secondary vertex mass distribution.

\begin{figure}[!t]
    \begin{center}
        \includegraphics[width=12.5cm]{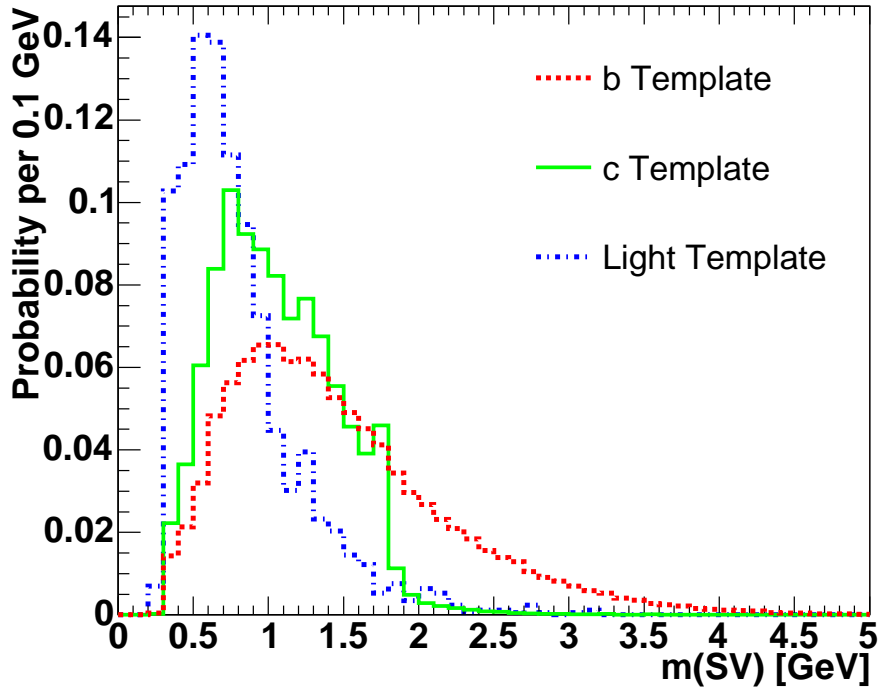}
        \caption{ \label{fig:TemplateComparison} Templates of the
        invariant mass of all tracks in a secondary vertex arising
        from bottom quarks, charm quarks, and light quarks and gluons from Monte Carlo simulations,
        normalized to unit area.  } \end{center}
\end{figure}

The fitting technique utilizes templates of the distribution of the
secondary vertex mass arising from the three primary sources expected
to contribute to the observed distribution: bottom quarks, charm
quarks, and light quarks or gluons.  These templates are obtained from
Monte Carlo samples containing final state photons~\cite{PhotonMC}.
The discriminating power of the secondary vertex mass is shown in
Fig.~\ref{fig:TemplateComparison}, in which the three templates are
normalized to unit area~\cite{charmPeak}.
The sum of the fractions of the three components is constrained to unity in the fit, $f_{b}+f_{c}+f_{\mathrm{light}}=1$.


This technique can be used to determine the number of events
containing a real photon and real heavy flavor in any sample.  We
first subtract the contribution due to misidentified photons by
applying the CES/CPR method to obtain the number of misidentified
photon events. We then estimate the fraction of heavy flavor in events
with a misidentified photon by fitting the secondary vertex mass
distribution in a sample enriched with jets faking photons, referred
to as the \emph{sideband} photon sample~\cite{sideband_photon}.  We
then subtract the number of events containing a misidentified photon
and heavy flavor from the number of events obtained from the standard
photon sample fit to obtain the number of $\gamma b$ and $\gamma c$
events.

In principle this technique could be directly applied to the search
region to obtain the number of $\gamma b $ events. However, this would
not give us the SM expectation, as the contribution from
any new process making such events would be counted~\cite{two_ways}.
Instead, the expected $\gamma b $ contribution is normalized by
applying this technique to a control region with a much larger SM
cross-section than that of the search region, and then extrapolating
to the search region by using efficiencies derived from the $\gamma+b$
Monte Carlo.  The final estimate for the number of $\gamma b$ events
in the search region is $N_{\gamma b}(\mathrm{search})=N_{\gamma
b}(\mathrm{control})\cdot\varepsilon(\mathrm{control}\rightarrow\mathrm{search})$.

We define the control region as the $\gamma+b$-tag sample, where the
only selection requirements are that there be at least one photon with
$|\eta|<1.1$ and $E_T>25$ GeV and one {\sc secvtx}-tagged jet having
$|\eta|<2$ and $E_T>15$ GeV.  The number of events in the search
region is less than $1\%$ of that in the control region, which
contains 93 894 events.
We obtain an efficiency of
$\varepsilon(\mathrm{control}\rightarrow\mathrm{search})=0.0123\pm0.0025$,
defined as the fraction of $\gamma+b$ Monte Carlo events in the
control region that survive the additional cuts of the search
region. The uncertainty on the efficiency is due to the differences in
jet multiplicities and \met\ distributions between data and the
background prediction in the control region.

\begin{figure}[htbp]
    \begin{minipage}{0.49\textwidth}
        \vspace{0pt}
     	\begin{center}
            \includegraphics[width=0.95\textwidth]{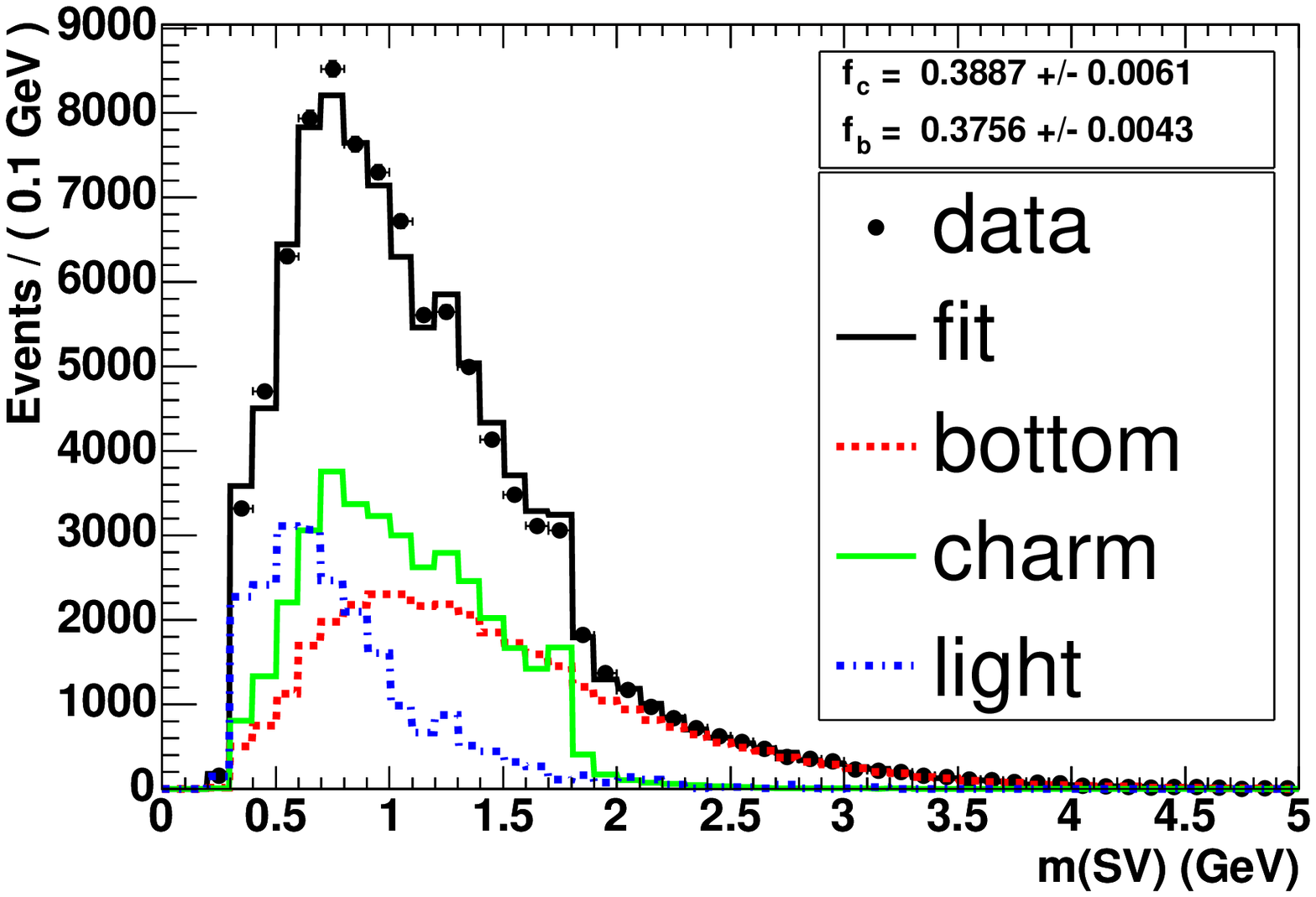}
       \end{center}
    \end{minipage}
    \hfill
    \begin{minipage}{0.49\textwidth}
       	\begin{center}
         	\vspace{25pt}
          \includegraphics[width=0.95\textwidth]{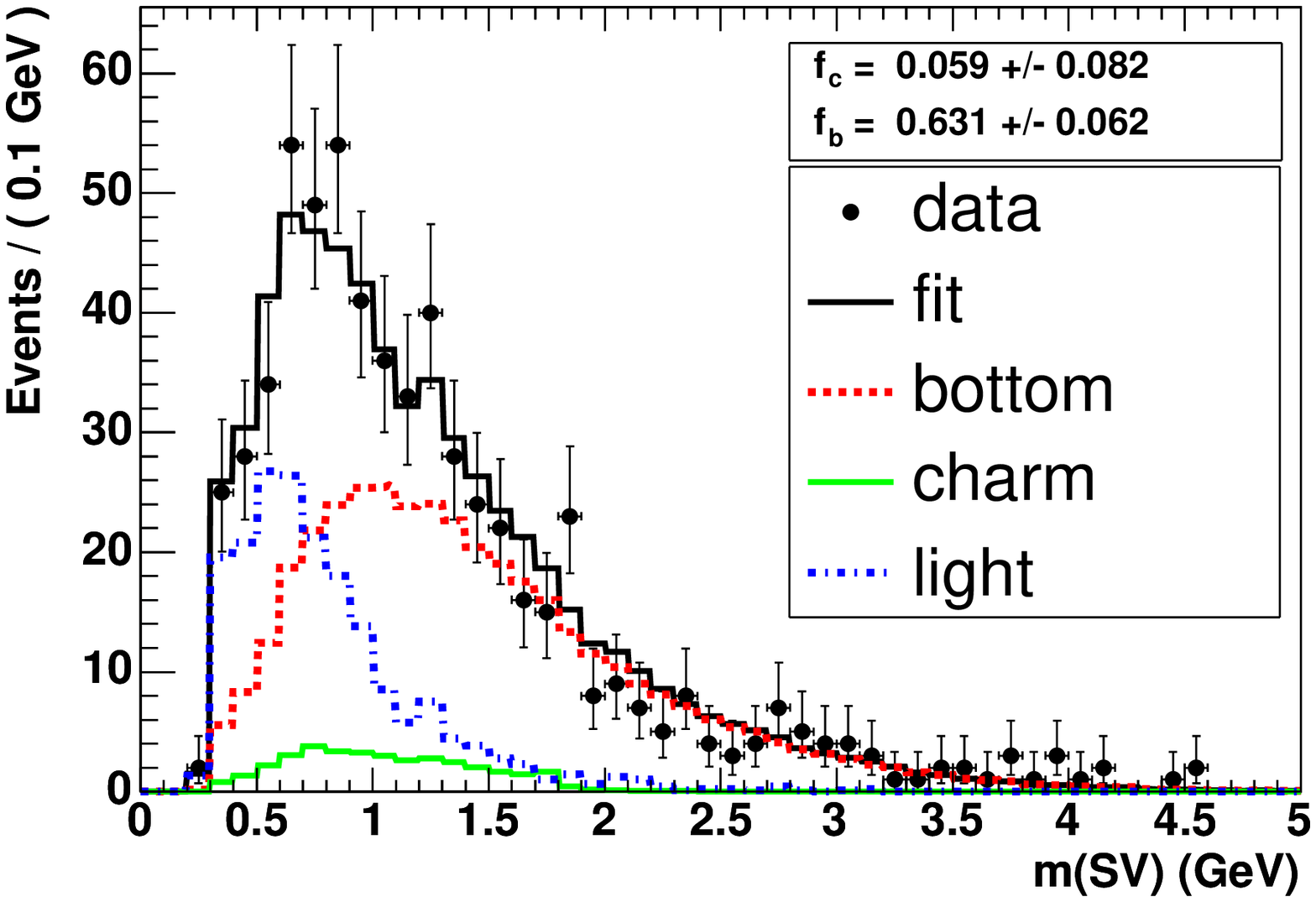}
        \end{center}`
    \end{minipage}
    \caption{\label{fig:mSVFit_ControlSignal_StdPho}
    The secondary vertex mass fit in events containing standard
        photons, for the control sample (left) and the search sample
        (right). Note the uncertainties on $f_c$ and $f_b$ are correlated 
	and purely statistical. } 

\end{figure}

Figure~\ref{fig:mSVFit_ControlSignal_StdPho} shows the results of a maximal likelihood fit performed
on the search and control region using the templates above to extract the fraction of $b$-jet and $c$-jet
events.  We estimate the
number of $\gamma b$ events by subtracting the misidentified photon
plus $b$ contribution from the control region and then multiplying by
$\varepsilon(\mathrm{control}\rightarrow\mathrm{search})$ to obtain
341 events with a statistical uncertainty of 18 events. The calculation of the systematic uncertainty on the
number of events is given in Sec.~\ref{sec:systErrs}.


The $\gamma c$ background is normalized by directly fitting the
secondary vertex mass in the search region.  We do not extrapolate the
charm normalization from the control sample because the uncertainties
on the matching scheme for charm quarks are large~\cite{zack} and
therefore the extrapolation efficiency would have large uncertainties.
After subtracting the misidentified photon plus charm contribution we
obtain an estimate of $9$ $\gamma c$ events with a statistical uncertainty of 52 events.

Note that because the charm background is measured in the search
region, this search is not sensitive to anomalous charm production.
It is, however, sensitive to anomalous production of the $\gamma
bj\met$ final state because we use $\gamma b$ Monte Carlo processes to
obtain the efficiency
$\varepsilon(\mathrm{control}\rightarrow\mathrm{search})$. 



\section{Systematic Uncertainties}
\label{sec:systErrs}
Sources of systematic uncertainty on the number of predicted events
arise from: a) the uncertainty on the true-photon weights $\wgamma$ in
the CES/CPR method, b) the uncertainty on the heavy-flavor mistag
prediction, and c) the uncertainty on the template shapes used in the
secondary vertex mass fit.  

The systematic uncertainty on \wgamma\ in the CES/CPR method arises
from uncertainties on the CES $\chi^2$ efficiencies and the CPR hit
rates for photons and backgrounds [$\epsilon_\mathrm{sig}$ and
$\epsilon_\mathrm{bkg}$ in Eq.~(\ref{eq:cescpr})]. The largest
uncertainty on the CES $\chi^2$ efficiencies is due to the gain
saturation in the CES detector~\cite{cescp2} (10\%\ on \wgamma).  The
largest uncertainty on the CPR hit rates is due to the modeling of the
hit rates of $\pi^0$ and $\eta^0$ (5--35\%\ on \wgamma).  On average,
$\wgamma$ has a relative systematic uncertainty of 11\%.

The uncertainty on the heavy-flavor mistag prediction comes from: the
finite size of data samples used for parameterization of the mistag
rates (10\%), variations between different data-taking periods
(6.5\%), and the uncertainty on a scale factor which takes into
account the contribution of misidentified $b$-tags from long-lived
hadrons ($\Lambda^0$, $K_s^0$) and secondary particles due to
interactions with detector material (10--15\%). More detail may be
found in Ref.~\cite{SECVTX}.

We estimate the systematic uncertainty arising from the secondary
vertex mass fitting procedure by varying the shapes of the templates
that are used in the binned likelihood fit.  The systematic effect of
mismodeled tracking inefficiency in the Monte Carlo simulation is estimated by
lowering the secondary vertex mass template mass scale by
$3\%$~\cite{cot_over_effic}.  We also refit the secondary vertex mass
distributions with templates derived from Monte Carlo samples that have the
$\met>25$ GeV cut imposed on them, as this may change the relative
fraction of semileptonic decays in the template samples and thereby
alter the secondary vertex mass distribution.  Because both of these
sources of uncertainty affect the shape of the templates, we take the
maximum variation observed as the systematic shift in normalization.
We obtain a $12\%$ uncertainty on the $b$-fraction and a $48\%$
uncertainty on the $c$-fraction from this estimate.

The numerical values of the systematic 
uncertainties are presented in Tab.~\ref{tab:BGresults} in
Sec.~\ref{sec:results} below. 
The CES/CPR method contributes $13\%$ of 
the systematic uncertainty on the total amount of background 
while the mistag parametrization and 
secondary vertex mass fit contribute $24\%$ and $63\%$ respectively.
The calculation of the total systematic uncertainty takes into account
correlations among the different sources of backgrounds to the
$\gbjmet$ signature. Because the CES/CPR method is used to estimate
the contribution of all four background categories defined in
Sec.~\ref{sec:bg}, we apply the CES/CPR systematic variations to all
backgrounds simultaneously when calculating the final CES/CPR
uncertainty on the total background prediction.  All other sources of
systematic uncertainty are combined as uncorrelated uncertainties.

\section{Results}
\label{sec:results}

We proceed to test the SM in the $\gbjmet$ signature in three ways:
comparing predicted event counts, looking for anomalous kinematic behavior, and counting additional objects in the events,
as might be expected from the production of new heavy states with extended decay chains. We also go beyond the Run I measurement
criteria by increasing the requirement on
missing transverse energy to 50 GeV, reducing the expected background contribution by a factor of $\approx 20$, and thereby
enhancing the sensitivity to new processes. The three tests are described in the sections below. 

\subsection{Comparing Predicted Event Counts}

Table~\ref{tab:BGresults} summarizes the background sources with associated
statistical and systematic uncertainties.  The total background prediction is
\begin{eqnarray}
\label{eq:totalBG}
N(\mathrm{BG})=607\pm74~(\mathrm{stat.})\pm86~(\mathrm{syst.}),
\end{eqnarray} where the first uncertainty is statistical and the
second systematic.
The observed number of events is 617, consistent with the background 
predictions.

\begin{table}[matching]
\caption{The numbers of predicted events from background sources. The
two uncertainties in each row are statistical and systematic, 
respectively. Note that the total systematic uncertainty is less than
the largest individual contribution due to an anti-correlation of the
CES/CPR uncertainties between the components.\label{tab:BGresults}}
\centering
\begin{tabular}{lccccc}
\hline
\hline
Background Source   & Expected Events &  & Stat &  & Sys \\
\hline
Misidentified $\gamma$ &115& $\pm$ &49& $\pm$ &54 \\
True $\gamma$, misidentified $b$ &141& $\pm$ &6& $\pm$ &30 \\
$\gamma b$  &341& $\pm$ &18& $\pm$ &91 \\
$\gamma c$  &9& $\pm$ &52& $\pm$ &14  \\
\hline
Total &607& $\pm$ &74& $\pm$ &86 \\
\hline
\hline
\end{tabular}
\end{table}


\subsection{Object Kinematics}
We examine three different types of distributions for anomalous shape
discrepancies with respect to the background prediction: the kinematics of
individual objects in the event such as jets and photons, global features of
the event such as $\met$, and the invariant masses of the combinations of objects.
%

\begin{figure}[htb]
    \begin{center}
        \includegraphics[width=0.75\textwidth]{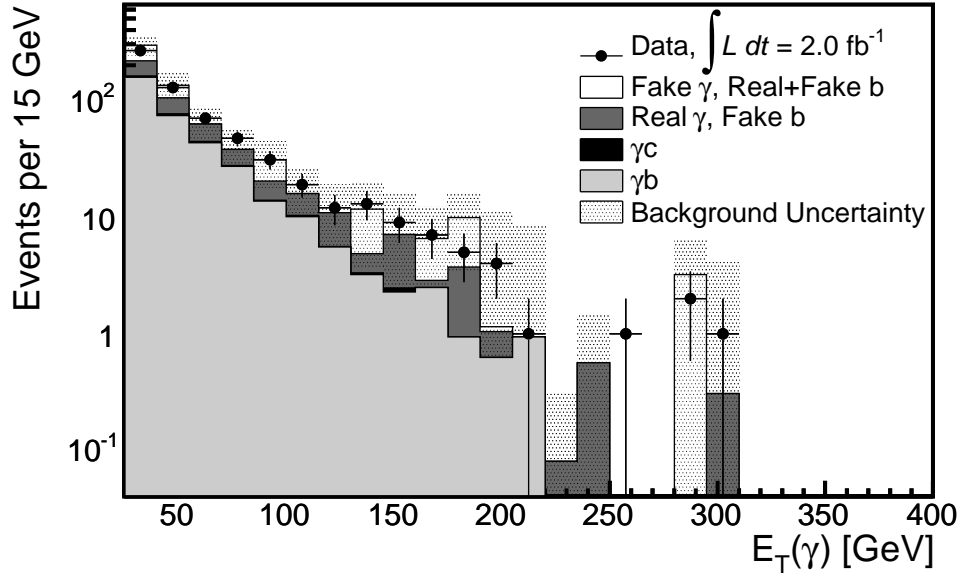}
            \caption{\label{fig:get}
        The distribution in photon $E_T$ observed (points) and from backgrounds (histogram).  The KS p-value is $63.7\%$.
    }
    \end{center}
\end{figure}

\begin{figure}[htbp]
  \begin{center}
    \includegraphics[width=0.75\textwidth]{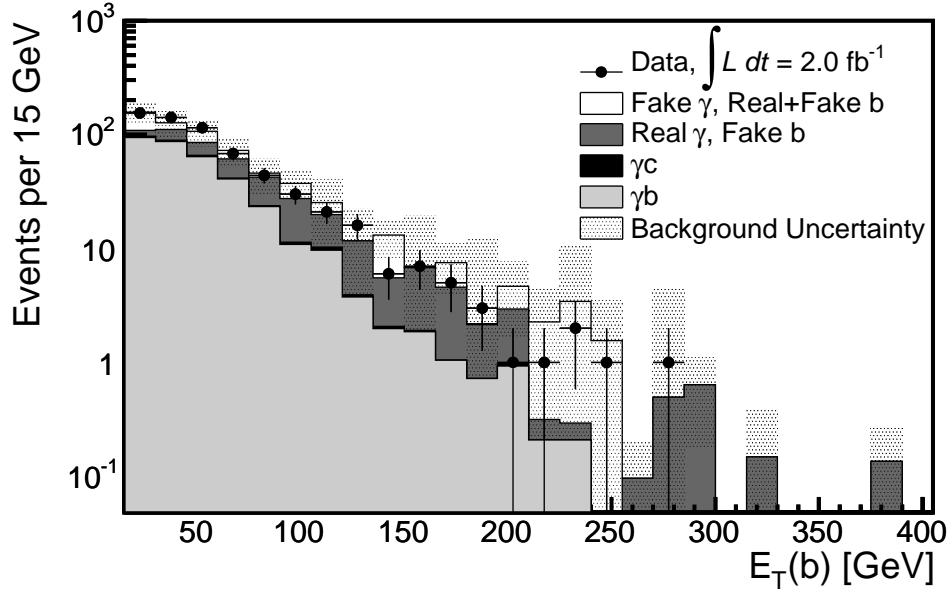}
    \caption{\label{fig:bet}
       The distribution in $b$-jet $E_T$ observed (points) and from backgrounds (histograms).  The KS p-value is $59.7\%$.
    }
    \end{center}
\end{figure}

\begin{figure}[htbp]
  \begin{center}
    \includegraphics[width=0.75\textwidth]{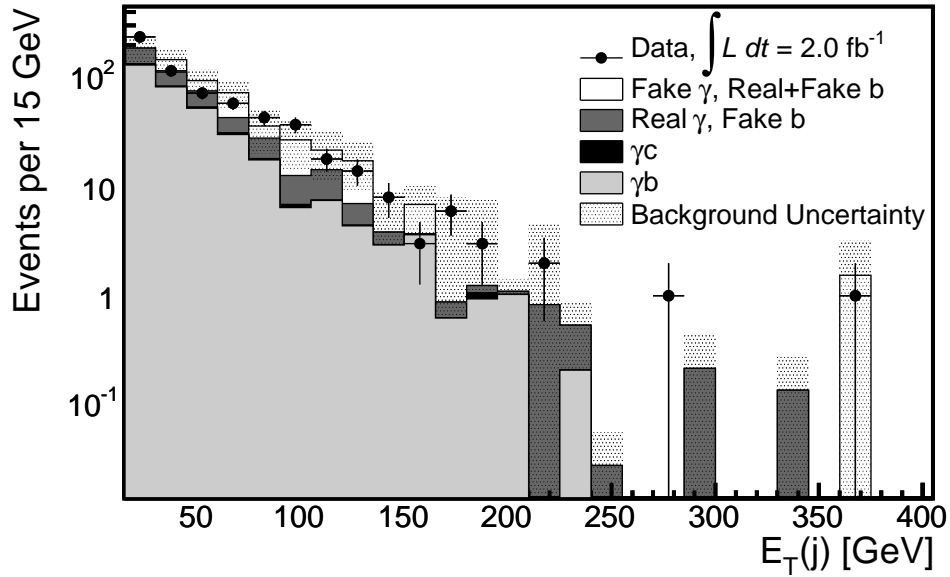}
    \caption{\label{fig:jet}
        The distribution in untagged jet $E_T$ observed (points) and from backgrounds (histogram).  The KS p-value is $10.4\%$.
}
    \end{center}
\end{figure}

\begin{figure}[htb]
  \begin{center}
    \includegraphics[width=0.75\textwidth]{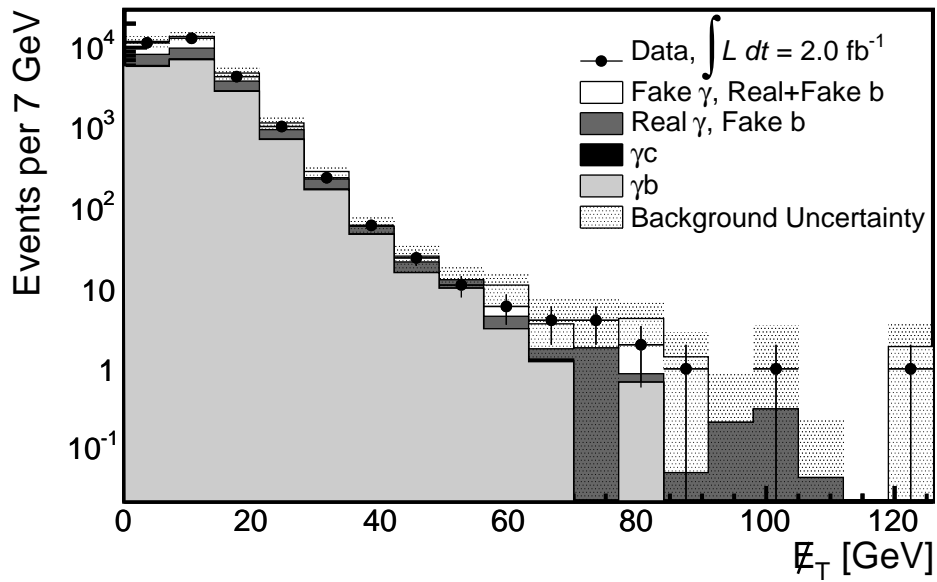}
    \caption{\label{fig:met}
        The distribution in missing transverse energy observed (points) and from backgrounds (histogram).  The KS p-value is $7.0\%$.}
    \end{center}
\end{figure}

\begin{figure}[htbp]
 \begin{center}
    \includegraphics[width=0.75\textwidth]{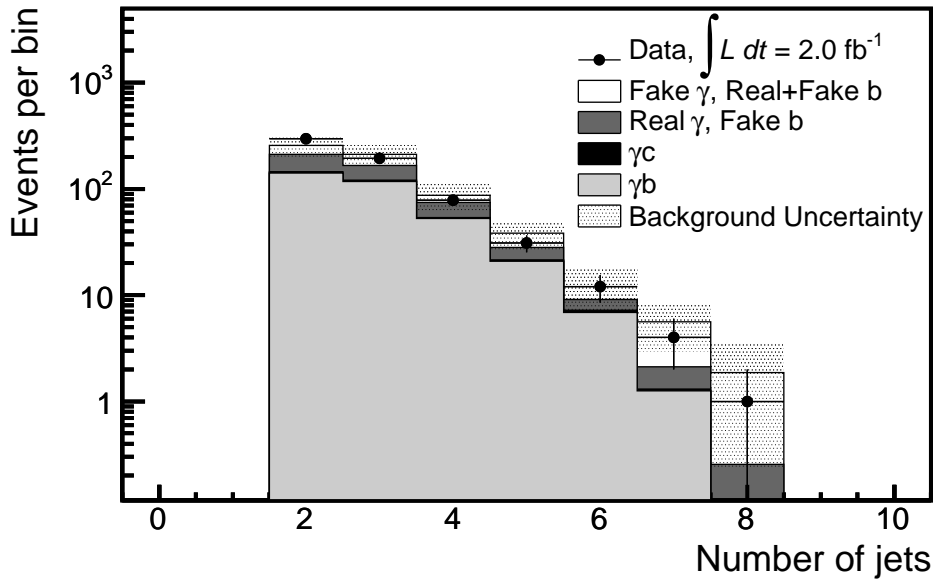}
    \caption{\label{fig:njet}
        The distribution in jet multiplicity observed (points) and from backgrounds (histogram) in logarithmic scale.  The KS p-value is $19.0\%$.}
    \end{center}
\end{figure}

\begin{figure}[htbp]
  \begin{center}
    \includegraphics[width=0.75\textwidth]{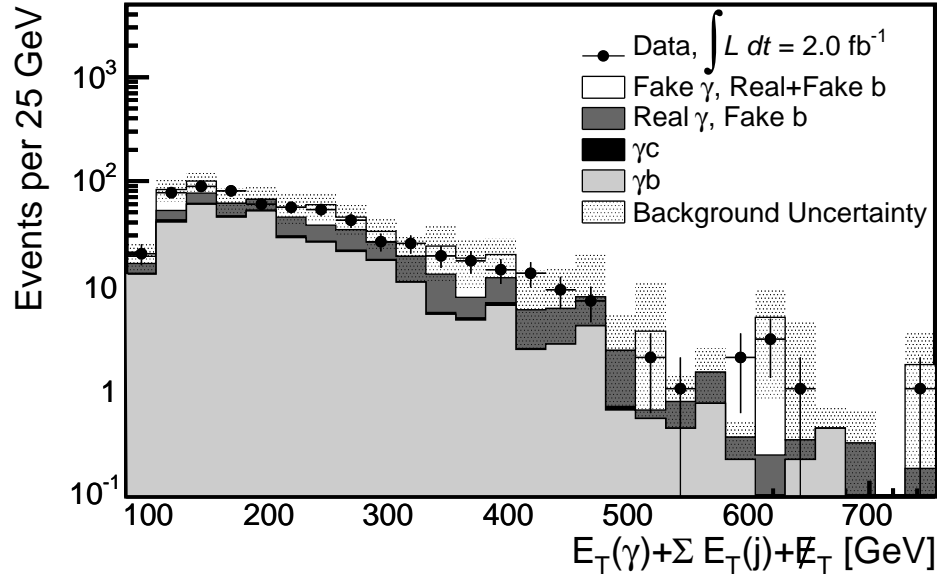}
    \caption{\label{fig:ht2}
        The distribution of the scalar sum of the transverse momenta of the $\gamma$, all jets in the event, and $\met$ observed (points) and from backgrounds (histogram).  The KS p-value is $99.7\%$.
        }
        \end{center}
\end{figure}

\begin{figure}[htb]
    \begin{center}
        \includegraphics[width=0.75\textwidth]{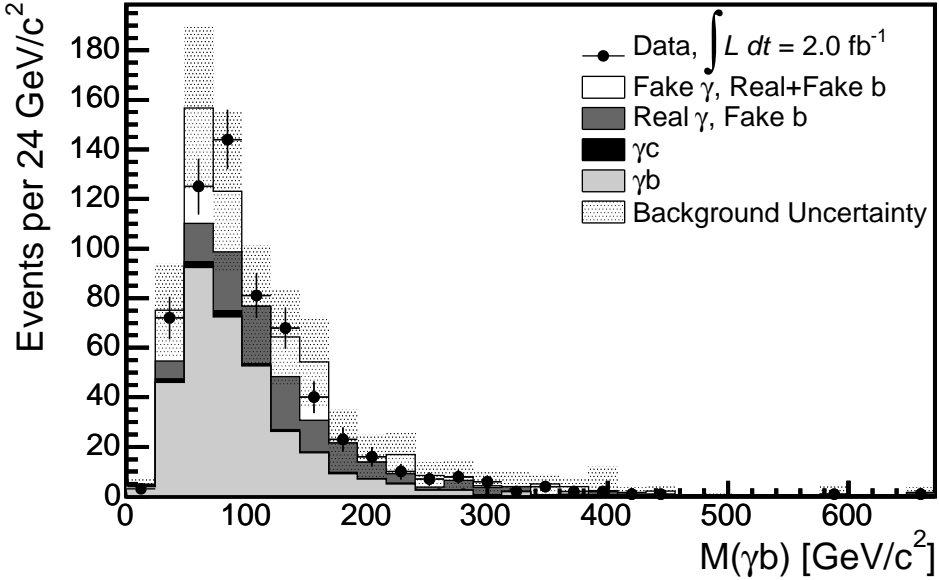}
        \caption{
            \label{fig:mgb}
            The distribution of the mass of the photon + $b$ jet observed (points) and from backgrounds (histogram).  The KS p-value is $62.0\%$.
            }
    \end{center}
\end{figure}

\begin{figure}[htb]
    \begin{center}
        \includegraphics[width=0.75\textwidth]{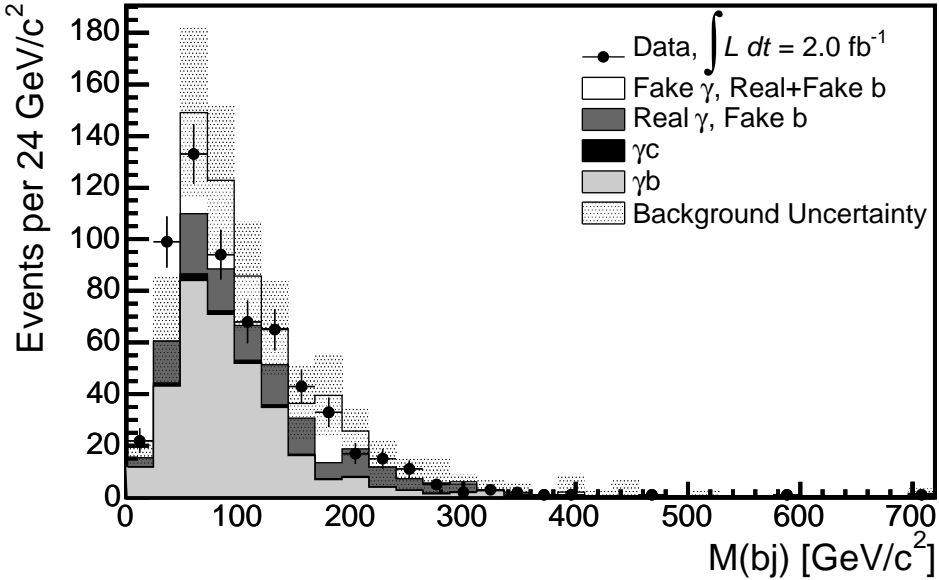}
        \caption{
            \label{fig:mbj}
            The distribution of the dijet mass observed (points) and from backgrounds (histogram).  The KS p-value is $9.8\%$.
            }
    \end{center}
\end{figure}

\begin{figure}[htb]
    \begin{center}
        \includegraphics[width=0.75\textwidth]{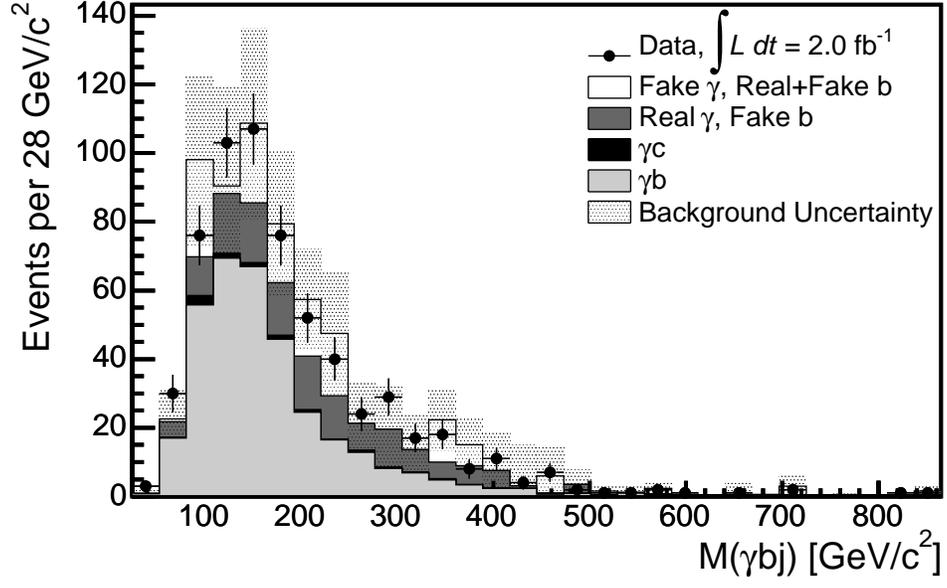}
        \caption{
            \label{fig:mgjj}
         The distribution of the invariant mass of the $\gamma$, $b$-jet, and $2^{nd}$ jet observed (points) and from backgrounds (histogram). 
The KS p-value is $99.8\%$.
         }
    \end{center}
\end{figure}

\begin{figure}[htb]
    \begin{center}
        \includegraphics[width=0.75\textwidth]{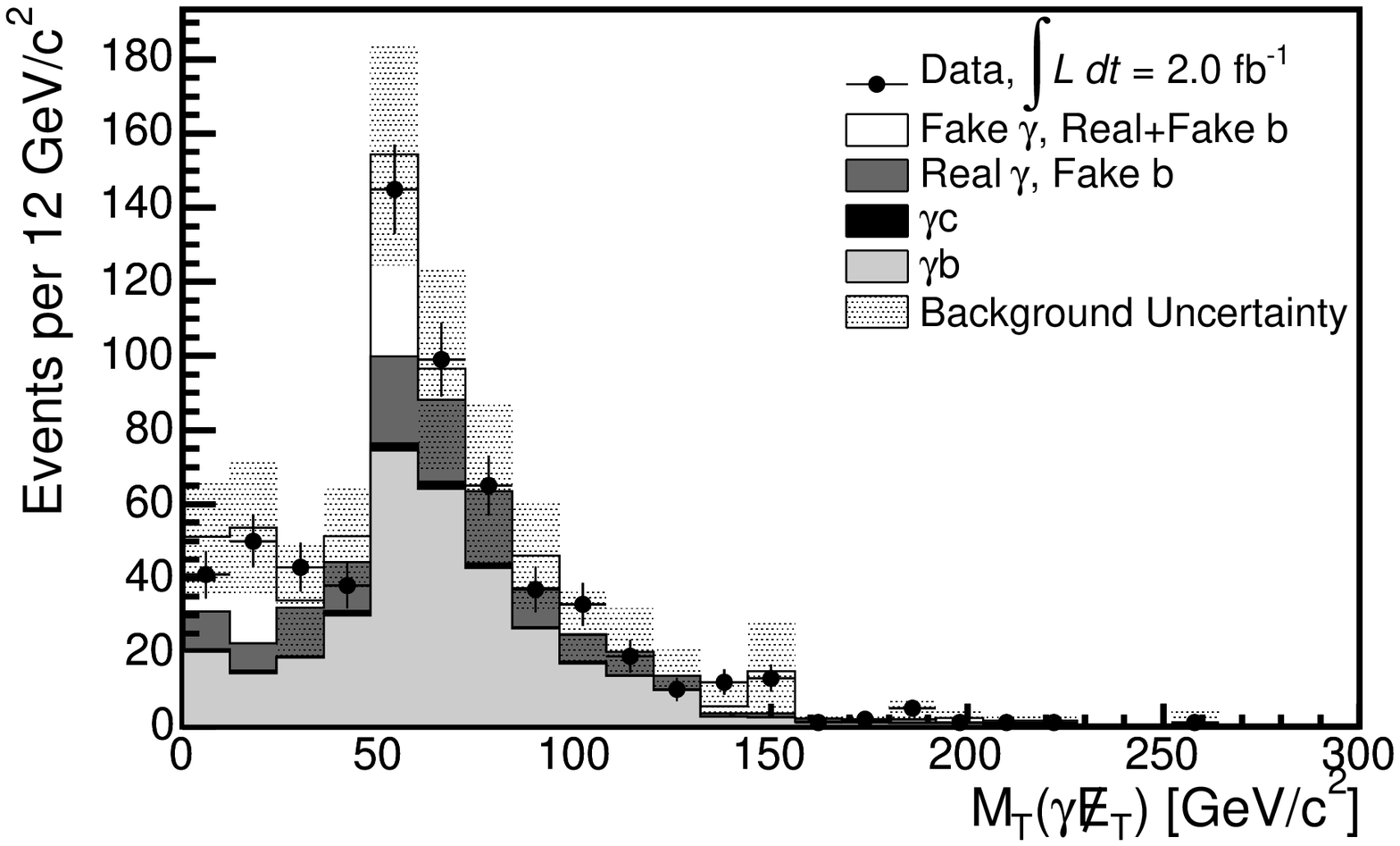}
        \caption{
        \label{fig:mgmet}
        The distribution of the transverse mass of the photon + $\met$ observed (points) and from backgrounds (histogram).  The KS p-value is $96.8\%$.}
    \end{center}
\end{figure}

\begin{figure}[htb]
    \begin{center}
        \includegraphics[width=0.75\textwidth]{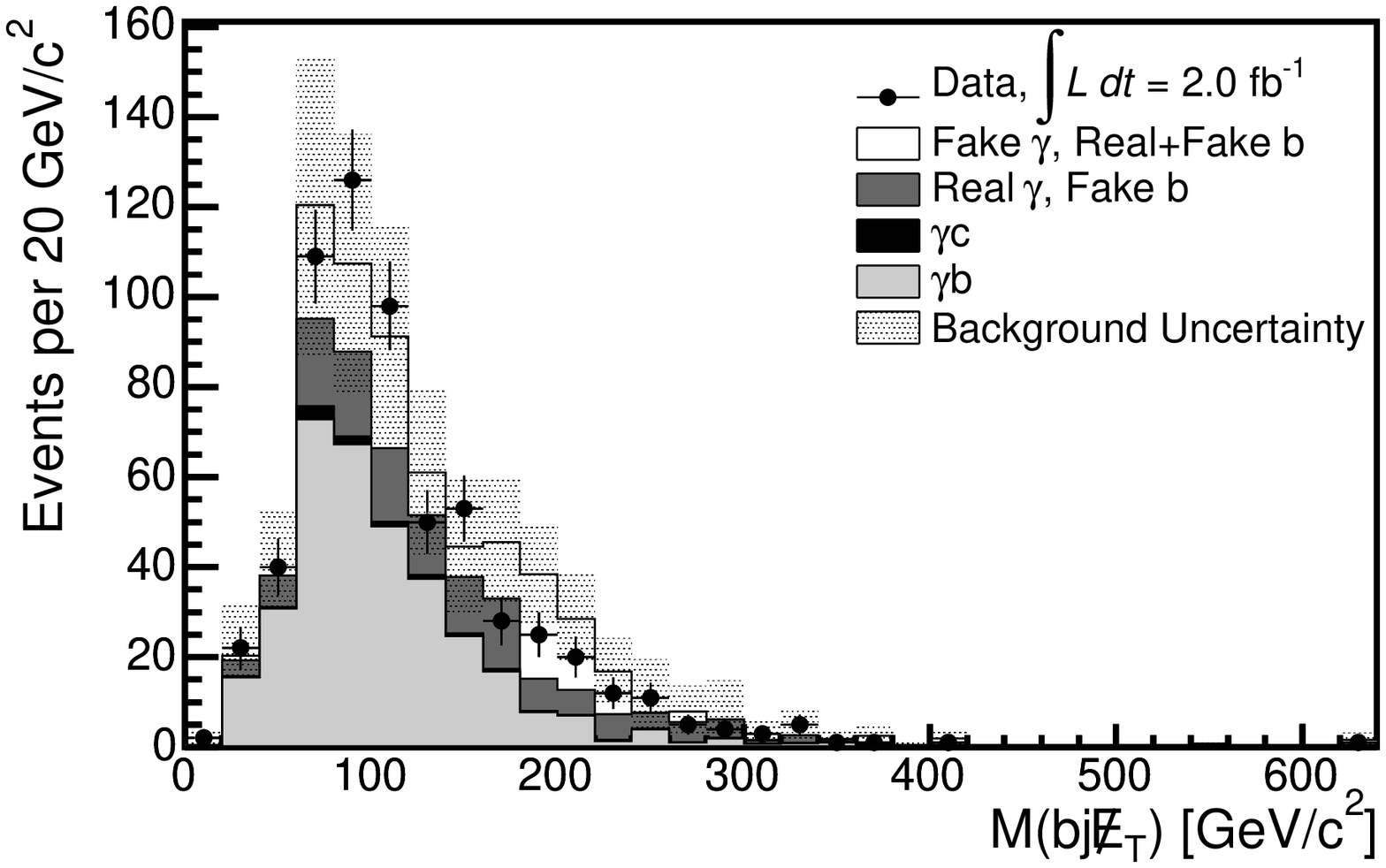}
        \caption{ \label{fig:mbjmet} The distribution of the
        transverse mass of the $b$-jet, $2^{nd}$ jet, and missing
        transverse energy observed (points) and from backgrounds
        (histogram).  The KS p-value is $21.1\%$.}  \end{center}
\end{figure}

The distributions of the transverse energy of the photon, $b$-jet, and $2^{nd}$ jet are shown in Figs.~\ref{fig:get},~\ref{fig:bet}, and~\ref{fig:jet} respectively.
%
The distributions of $\met$, $N(\mathrm{jets})$, and $H_T$, where $H_T$ is the scalar sum of the transverse momenta of the photon, all jets in the event, and $\met$ are shown in Figs.~\ref{fig:met},~\ref{fig:njet}, and~\ref{fig:ht2} respectively.  The $\met$ distribution is shown before the application of the $\met>25$ GeV cut but after the application of all other selections.
The distributions of $M(\gamma b)$, $M(bj)$, $M(\gamma bj)$,
$M_T(\gamma\met)$, and $M_T(bj\met)$ are shown in
Figs.~\ref{fig:mgb},~\ref{fig:mbj},~\ref{fig:mgjj},~\ref{fig:mgmet},
and~\ref{fig:mbjmet} respectively.  The transverse mass $M_T$ is
calculated with the transverse components of object momenta:
\begin{eqnarray}
  M_T = \sqrt{\left(\Sigma_i E_T^i\right)^2 - \left(\Sigma_i p_x^i\right)^2 - \left(\Sigma_i p_y^i\right)^2},
\end{eqnarray}
 where $E_T^i$, $p_x^i$, and $p_y^i$ are the transverse energy and $x$ and $y$
components of the momentum of object $i$ (which could be a
photon, $b$-quark jet, jet, or missing energy).
Note that the binning for all distributions is such that there are
no overflows.

We test the consistency between the observed shapes of kinematic
distributions and the shape predicted by the background expectation by
running pseudoexperiments for each distribution studied and
calculating the Kolmogorov-Smirnov (KS) distance for each
pseudoexperiment.  The probability that a random sampling of the
estimated background distribution would give a higher KS distance than
the observed data distribution, referred to as the ``KS p-value'', is
obtained for each kinematic variable studied by integrating the tail of
the distribution of KS distances.  We obtain a range of KS p-values
between $7.0\%$ and $99.8\%$, indicating that the kinematic
distributions observed are consistent with background expectations.

\subsection{Effect of Additional Selections}
\label{sec:morecuts}
We further investigate the existence of possible anomalies in the
$\gamma bj\met X$ final state by making additional selections and
comparing the number of observed events to the background predictions.
We chose criteria based on expected SM distributions and selections
used previously in the search of Ref~\cite{RunI_gammabjmet}.  The additional
selections we make are $\met>50$ GeV, $N(\mathrm{jets})\geq 3$,
$p_{T}(\gamma)>50$ GeV, $H_{T}>200$ GeV, $E_{T}(b)>50$ GeV, and
$\Delta\phi(\mathrm{jet},\met)>0.5$.

Table~\ref{tab:additionalselect} summarizes the effects of the
additional selections.  We apply these in two different ways: one at a
time independently of all other additional selections, and after the
application of the $\met>50$ GeV selection.  No anomalous excess of
events is observed.

\begin{figure}[htb]
    \begin{center}
        \includegraphics[width=0.75\textwidth]{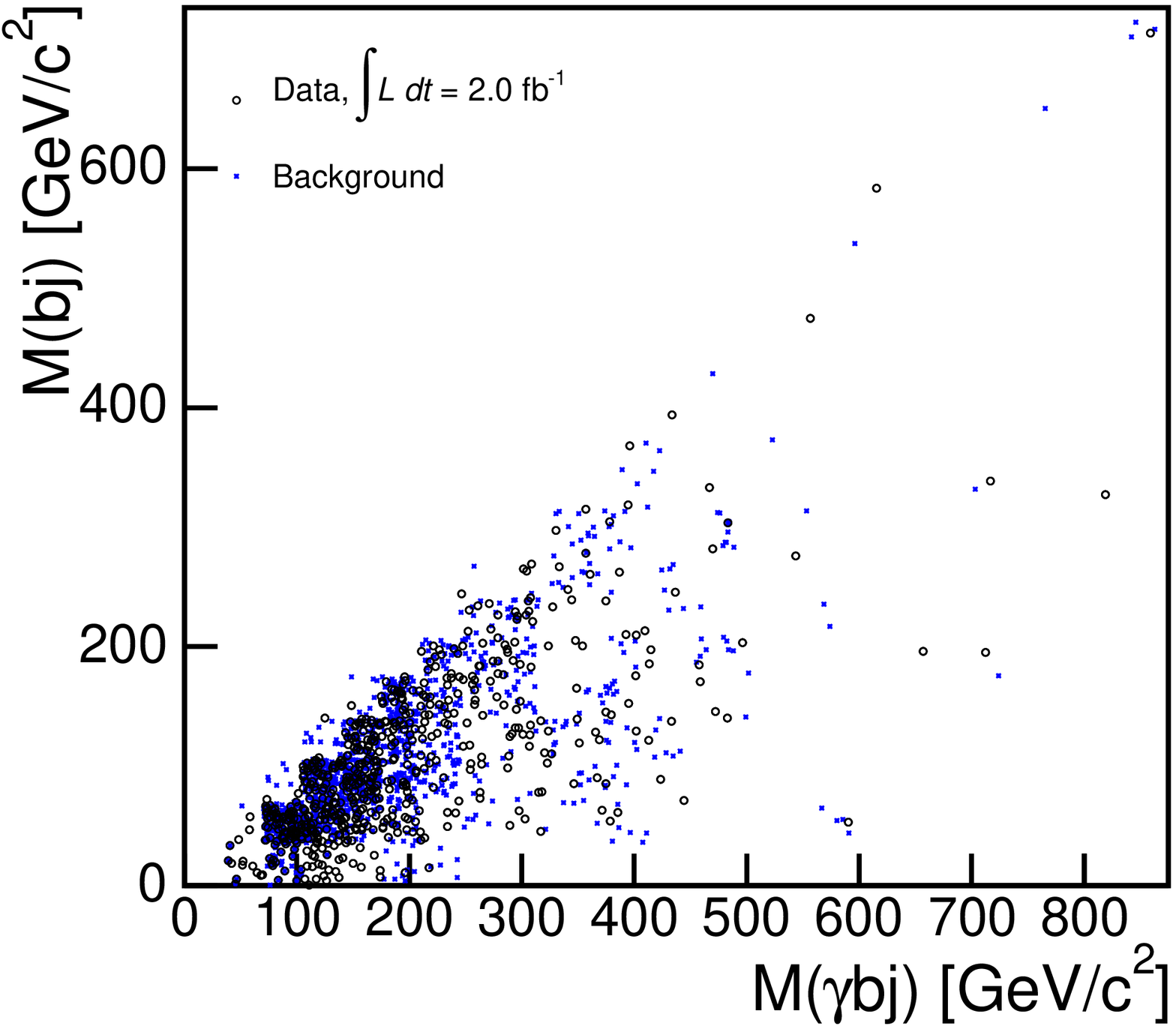}
        \caption{
        \label{fig:mgjjvsmjj}
	$M(bj)$ versus $M(\gamma bj)$ for the events which satisfy 
	the selections in Tab.~\ref{tab:evselect}, observed (big dots) 
	and expected (small dots). 
        }
    \end{center} 
\end{figure}

Finally, exotic particles with cascade decays, 
$X \rightarrow \gamma Y \rightarrow bj$, may form a cluster in the scatter 
plot of $M(\gamma bj)$ vs. $M(bj)$. In Fig.~\ref{fig:mgjjvsmjj}, we compare the
 observed distribution to that from the estimated background; we do not 
see any evidence of an anomaly.

\begin{table}[htb]
\caption{The number of events observed and the 
predicted background for additional independent selections.  The first
uncertainty in the observed columns is statistical and the second is
systematic.
\label{tab:additionalselect}}
\begin{center}
\begin{tabular}{lcccccc  cccccc}
\hline
\hline
\textbf{Selection} & \multicolumn{6}{c}{\textbf{No additional cuts}} & \multicolumn{6}{c}{\textbf{With $\met>50$ GeV}} \\
\hline
& Observed & \multicolumn{5}{c}{Predicted}& 
Observed & \multicolumn{5}{c}{Predicted} \\
\hline
$\met>50$ GeV                       & 28  &30& $\pm$ &10& $\pm$ &5  &  &  &       & &       & \\				  
$N(\mathrm{jets})\geq 3$            & 321 &329& $\pm$ &46& $\pm$ &46&15&17& $\pm$ &7& $\pm$ &3\\	  
$p_{T}(\gamma)>50$ GeV              & 257 &247& $\pm$ &42& $\pm$ &39&16&21& $\pm$ &8& $\pm$ &5\\ 	  
$H_{T}>200$ GeV                     & 304 &322& $\pm$ &45& $\pm$ &46&25&28& $\pm$ &9& $\pm$ &5\\	  
$E_{T}(b)>50$ GeV                   & 286 &310& $\pm$ &43& $\pm$ &44&18&22& $\pm$ &8& $\pm$ &6\\	  
$\Delta\phi(\mathrm{jet},\met)>0.5$ & 343 &368& $\pm$ &47& $\pm$ &49&15&16& $\pm$ &8& $\pm$ &4\\	  
\hline
\hline
\end{tabular}
\end{center}
\end{table}

\section{Conclusions}
\label{sec:conclusions}
We have searched for the anomalous production of events containing a
 photon, two jets, one which is identified as originating from a $b$
 quark, and missing transverse energy.  The number of events observed
 in data is consistent with the number of expected background
 events. No significant excess of events with respect to the
 background prediction is observed in any of the kinematic
 distributions studied.  The shapes of these distributions are
 consistent with SM expectations.  Furthermore, we do not see any
 anomalous production of events after applying additional selections.
We conclude that the $2.0\pm0.1~\mathrm{fb}^{-1}$ $\gamma+b+j+\met+X$ sample is
consistent with SM background expectations.  

\begin{acknowledgments}
We thank Stephen Mrenna for teaching us how to implement the matching of matrix element n-jet channels.
We also would like to thank Zack Sullivan for helpful discussions and Michel Herquet and Johann Alwall
for providing resources and support for {\sc madgraph}.

We thank the Fermilab staff and the technical staffs of the participating institutions for their vital contributions. This work was supported by the U.S. Department of Energy and National Science Foundation; the Italian Istituto Nazionale di Fisica Nucleare; the Ministry of Education, Culture, Sports, Science and Technology of Japan; the Natural Sciences and Engineering Research Council of Canada; the National Science Council of the Republic of China; the Swiss National Science Foundation; the A.P. Sloan Foundation; the Bundesministerium f\"ur Bildung und Forschung, Germany; the Korean Science and Engineering Foundation and the Korean Research Foundation; the Science and Technology Facilities Council and the Royal Society, UK; the Institut National de Physique Nucleaire et Physique des Particules/CNRS; the Russian Foundation for Basic Research; the Ministerio de Ciencia e Innovaci\'{o}n, and Programa Consolider-Ingenio 2010, Spain; the Slovak R\&D Agency; and the Academy of Finland. 
\end{acknowledgments}


\begin{thebibliography}{99}
\bibitem{EtPt}
The CDF~II detector uses a cylindrical coordinate system in which $\phi$ is the
azimuthal angle, $r$ is the radius from the nominal beam line, and $z$
points in the proton beam direction.
The transverse ($r-\phi$) plane is perpendicular to the $z$
axis. Transverse momentum and energy are the respective 
projections of momentum
measured in the tracking system and energy measured in the calorimeter
system onto the $r-\phi$ plane, and are defined as $p_T = p\sin\theta$
and $E_T = E\sin\theta$.  Here, $\theta$ is the polar
angle measured with respect to the interaction vertex.
%
%
Missing $\rm E_T$ ($\metvec$) is defined by $\metvec = -\sum_{i} E_T^i
\hat{n}_i$, where $i$ is the calorimeter tower number for $|\eta| <$ 3.6, and
$\hat{n}_i$ is a unit vector perpendicular to the beam axis and pointing at
the $i^\mathrm{th}$ tower.
The pseudorapidity $\eta$ is defined as $-\ln(\tan(\theta/2))$, where
$\theta$ is measured with respect to the origin of the detector.
We define the magnitude $\met=|\metvec|$.
%
We use the convention that ``momentum'' refers to $pc$ and ``mass'' to
$mc^2$.
\bibitem{Photon25Trig}
  The trigger requires one electromagnetic (EM) 
cluster with (i) $E_T > $ 25~GeV, and (ii) 
isolation energy less than 2~GeV or $0.1\times E_T$. 
The energy of the EM cluster is the sum of the energies measured  
in two adjacent calorimeter towers at the same wedge, i.e. 
$\Delta \phi \times \Delta \eta \approx 15^{\deg} \times 0.2$. 
The isolation energy is calculated as the sum of $E_T$ of all towers (except 
for the EM cluster) in the cone with radius $R=0.4$ centered around the 
EM cluster.
\bibitem{photonHADleakage}
    The ratio of hadronic energy to electromagnetic energy,
$E_{\mathrm{had}}/E_{\mathrm{em}}$, must be less than
$0.055+0.00045\times E_{em}(\mathrm{GeV})$.
\bibitem{isodef}
	Tracks with $p_T>500$ MeV are included in the isolation sum,
$\sum_{i}p_{T}^{i}$, if they are in a cone of size $\Delta R=0.4$
around the isolated track candidate, have greater than $23$ ($19$)
axial (stereo) hits in the tracking chamber, and have beam-corrected
track impact parameter $|d_{0}|<250$ $\mu$m. In addition, they are
required to have $\geq3$ silicon hits on the track if the number
expected is $\geq3$, and a difference of the $z$ component of closest
approach to the primary vertex of $|z_{i}-z_{\mathrm{track}}|<5$ cm
with $|z_{track}|<60$ cm.
\bibitem{mistag}The heavy-flavor mistag rate is the probability that a
  light quark or gluon jet will be misidentified as heavy flavor by
  the {\sc secvtx} algorithm.  This rate is of order $1\%$.
\bibitem{eiko_endnote} The values of $\epsilon_\mathrm{sig}$ 
and $\epsilon_\mathrm{bkg}$ are first determined from the standard CDF 
simulation of single photons and a mixture of mesons 
according to their relative productions and reconstruction efficiencies, 
where the CES and CPR responses are tuned based on electron test beam
and $W$ leptonic decays.  We then calibrate $\epsilon_\mathrm{sig}$
with a pure photon sample obtained from $W^-\rightarrow \ell^-
\bar{\nu}_{\ell}\gamma$ and $Z^0\rightarrow
\ell^+\ell^-\gamma$ decays. The efficiency for background, 
$\epsilon_\mathrm{bkg}$, is calibrated using the isolation
distribution of inclusive photon candidates.
\bibitem{PhotonMC} Because the event topology of photon events may be different from that of
generic multi-jet events, we use templates that are derived from
samples containing true photons.  The bottom and charm templates are
obtained from {\sc madgraph} samples as described in Section~\ref{sec:bg} 
and the light template is
obtained from a {\sc pythia} inclusive photon sample with the heavy
flavor removed.
\bibitem{charmPeak} The feature in the charm template near $m(SV)=1.8$ GeV/c$^{2}$ is attributable to $D^0$ and $D^{+}$ decays where the invariant mass of the constituent tracks almost reproduces the mass of the decaying hadron.  This peak is prominent because the $D^{0}$ and $D^{+}$ hadrons are the most common components of charm jets.
\bibitem{sideband_photon}
The sideband-photon candidates satisfy loose identification requirements
but fail the standard requirements as described in Ref.~\cite{diphoton_MET}. 
The loose photon identification has reduced requirements on 
the shower profile, leakage into the hadron calorimeter, and isolation. 
It has been estimated that only $\approx 5\%$ of the sideband-photon sample 
has true photons.
\bibitem{two_ways} Applying this secondary mass fit technique to the
  search sample gives $325\pm40$ $\gamma b$ events, compared to the
  prediction obtained directly from the control sample of $341\pm18$ events,
  where the uncertainties are statistical only.
\bibitem{cot_over_effic} This is derived from comparisons of tracking
efficiency between Monte Carlo simulation and data. The effect of
possible mismodeling is then propagated to the mass of secondary vertices.
\end{thebibliography}
\end{document}